\documentclass[aps,prd,preprintnumbers,nofootinbib,twocolumn]{revtex4}
\usepackage{bm}
\usepackage{latexsym}
\usepackage{dcolumn}
\usepackage{amsmath,amsfonts,amssymb}
\usepackage{graphicx,epsfig}
\usepackage{psfrag}
\usepackage{amsthm}

\def\be {\begin{equation}}
\def\ee {\end{equation}}
\def\bea {\begin{eqnarray}}
\def\eea {\end{eqnarray}}
\def\bc {\begin{center}}
\def\ec {\end{center}}
\def\bfg {\begin{figure}}
\def\efg {\end{figure}}
\def\bi {\begin{itemize}}
\def\ei {\end{itemize}}
\def\nn {\nonumber}

\def\la {\label}
\def\le {\left}
\def\ri {\right}
\def\pa {\partial}

\def\no {\noindent}

\def\vs {\vspace}

\def\a  {\alpha}
\def\b  {\beta}

\def\d  {\delta}
\def\D  {\Delta}

\def\beq{\begin{equation}}
\def\eeq{\end{equation}}
\def\br{\begin{eqnarray}}
\def\er{\end{eqnarray}}
\newcommand{\eel}[1] {\label{#1}\end{equation}}

\newcommand{\bdm}{\begin{displaymath}}
\newcommand{\edm}{\end{displaymath}}

\begin{document}
\title{Planck scale effects on some low energy quantum phenomena}

\author{Saurya Das} \email[email: ]{saurya.das@uleth.ca}

\affiliation{
Theoretical Physics Group,
Dept. of Physics and Astronomy,
University of Lethbridge, 4401 University Drive,
Lethbridge, Alberta, Canada T1K 3M4}

\author{R. B. Mann}\email[email:]{rbmann@sciborg.uwaterloo.ca}

\affiliation{
Dept. of Physics, University of Waterloo, 200 University Avenue West,
Waterloo, Ontario, Canada N2L 3G1
}

\begin{abstract}
Almost all theories of Quantum Gravity predict modifications of the Heisenberg Uncertainty Principle
near the Planck scale to a so-called Generalized Uncertainty Principle (GUP). Recently
it was shown that the GUP gives rise to corrections to the Schr\"odinger and Dirac equations, which in turn affect
all non-relativistic and relativistic
quantum Hamiltonians. In this paper, we apply it to superconductivity and the quantum Hall effect
and compute Planck scale corrections. We also show that Planck scale effects may account for
a (small) part of the anomalous magnetic moment of the muon.
%
%
We obtain (weak) empirical bounds on the undetermined GUP parameter  from
present-day experiments.
\end{abstract}

\maketitle


Various approaches to quantum gravity,
such as String Theory, Doubly Special Relativity (DSR) Theories, Loop Quantum Gravity via so-called
Polymer Quantization, as well as black hole physics, predict a minimum measurable length, and a
modification of the Heisenberg Uncertainty Principle to a so-called
Generalized Uncertainty Principle, or GUP, and a corresponding modification of the
commutation relations between position and momenta
\cite{guppapers,kmm,kempf,brau,sm,cg,viqar}. The only GUP
consistent with the symmetries and index structure of the
modified commutator bracket between position and momentum from all the above
derivations (all of which predict corrections involving at most terms up to second order in the momentum),
{\it and}
which ensures $[x_i,x_j]=0=[p_i,p_j]$ (via the Jacobi identity)\footnote{
(a) In refs. \cite{advplb1,dvprl,advprd}   $\a$ was used in place of $a$.  \\
(b) The results of this article do not depend on this particular form of GUP chosen,
and continue to hold for a a large class of variants,
so long as an ${\cal O}(a)$ term is present in the right hand side
of Eq.(\ref{comm01}).
} is, to the best of our knowledge \cite{advplb1,advprd}
\bea
[x_i, p_j]\hspace{-1ex} &=&\hspace{-1ex} i \hbar\hspace{-0.5ex} \left[  \delta_{ij}\hspace{-0.5ex}
- \hspace{-0.5ex} a \hspace{-0.5ex}  \le( p \delta_{ij} +
\frac{p_i p_j}{p} \ri)
+ a^2 \hspace{-0.5ex}
\le( p^2 \delta_{ij}  + 3 p_{i} p_{j} \ri) \hspace{-0.5ex} \ri]
\label{comm01} \\
%
%
%
 \Delta x \D p \hspace{-1ex}&\geq &\hspace{-1ex}\frac{\hbar}{2}
\le[ 1 - 2 a <p> + 4a^2 <p^2>
\ri] ~ \nn \\
\hspace{-1ex}&\geq& \hspace{-1ex}
\frac{\hbar}{2} \hspace{-1ex}
\le[\hspace{-0.5ex} 1\hspace{-0.5ex}  +\hspace{-0.7ex}  \le(\hspace{-0.7ex}  \frac{a}{\sqrt{\langle p^2 \rangle}} +\hspace{-0.2ex}4a^2 \hspace{-0.9ex} \ri)
\hspace{-0.6ex}  \D p^2 \hspace{-0.6ex}
+\hspace{-0.6ex}  4a^2 \langle p \rangle^2 \hspace{-0.6ex}
- \hspace{-0.6ex}  2a \sqrt{\langle p^2 \rangle}\hspace{-0.2ex}
\ri]\hspace{2ex} \label{dxdp1}
\eea
where
$a = {a_0}/{M_{Pl}c} = {a_0 \ell_{Pl}}/{\hbar},$
$M_{Pl}=$ Planck mass, $\ell_{Pl}\approx 10^{-35}~m=$ Planck length,
and $M_{Pl} c^2=$ Planck energy $\approx 10^{19}~GeV$.
It should be stressed that the GUP-induced terms become important near the Planck scale.
It is normally assumed that $a_0 \approx 1$.
%
For phenomenological implications of the above GUP, see
\cite{dvprl,dvcjp,advplb1,advprd,advplb2}.
Note that although Eqs. (\ref{comm01}) and (\ref{dxdp1}) are not
Lorentz covariant, they are at least approximately covariant under
DSR transformations \cite{cg}. We expect the results of our paper
to have similar covariance as well. In addition, since DSR transformations
preserve not only the speed of light, but also the Planck momentum
and the Planck length, it is not surprising that Eqs.
(\ref{comm01}) and (\ref{dxdp1}) imply the following minimum
measurable length {\it and} maximum measurable momentum
\bea
\D x &\geq& (\D x)_{min}  \approx a_0\ell_{Pl} \la{dxmin} \\
\D p &\leq& (\D p)_{max} \approx \frac{M_{Pl}c}{a_0}~. \la{dpmax}
\eea
\par\noindent
It can be shown that the following definitions
\bea x_i = x_{0i}~,~~
p_i = p_{0i} \le( 1 - a p_0 + 2a^2 p_0^2 \ri)~, \la{mom1}
\eea
(with $x_{0i}, p_{0j}$
satisfying the canonical commutation relations
$ [x_{0i}, p_{0j}] = i \hbar~\delta_{ij}, $
such that $p_{0i} = -i\hbar \partial/\partial{x_{0i}}$)
satisfy Eq.(\ref{comm01}).
In \cite{dvprl,advplb1} it was shown using Eq.(\ref{mom1}), that
any non-relativistic Hamiltonian of the form
$H=p^2/2m + V(\vec r)$ can be written as $H = p_0^2/2m - (a/m)p_0^3 + (5a^2/2m)p_0^4
+ V(r) + {\cal O}(a^3)$, implying the modified Schr\"odinger equation
\bea
\le[
-\frac{\hbar^2}{2m} \nabla^2 + \frac{i\a \hbar^3}{m} \nabla^3 + \frac{5a^2\hbar^4}{2m} \nabla^4
\ri]\psi = i\hbar \frac{\partial \psi}{\partial t}~.
\la{schr1}
\eea
We will treat the $a$ and $a^2$ terms as perturbations, although
the higher order Schr\"odinger equation now has new {\it non-perturbative}
solutions of the form $\psi \sim e^{ix/2a\hbar}$, which may have interesting physical implications \cite{advplb1}.
Some phenomenological implications of the above GUP modified Hamiltonian were examined
in \cite{dvcjp,advplb2}.

Note that for the earlier versions of the GUP (which did not take into account DSR), the
terms in Eqs.(\ref{comm01}), (\ref{dxdp1}) and (\ref{mom1}) linear in $a$ and the Planck length were
effectively absent. In the following sections, we will apply that version to the problems of Superconductivity
(Section \ref{superconductivity}) and the Quantum Hall Effect (Section \ref{quantumhall}).
In Section \ref{muon}, we will write the Dirac equation that follows from
the full GUP, and apply it to the problem of anomalous magnetic moment
of the muon.

\section{Superconductivity}
\la{superconductivity}

The usual Schr\"odinger current minimally coupled to a Cooper pair of charge $-2e$ and
mass $2m$ reads \cite{am}
%
\be
\vec J = -\frac{e}{2m} \le[ \psi^\star
\le\{\le( \frac{\hbar}{i}\vec\nabla + \frac{2e}{c}\vec A \ri) \psi\ri\}
+ \le\{\le( \frac{\hbar}{i}\vec\nabla + \frac{2e}{c}\vec A \ri) \psi\ri\}^\star \psi \ri]~.
\ee
Substituting $\psi = |\psi|e^{i\phi}$, and assuming virtually all
spatial dependence of the wavefunction in the phase $\phi$, such that
$|\psi| \approx $  constant, and $\vec\nabla\psi=i\psi\vec\nabla\phi$, we get:
\be
\vec J = -\le[
\frac{2e^2}{mc}\vec A + \frac{e\hbar}{m} \vec\nabla\phi~.
\ri] |\psi|^2
\la{current2}
\ee
%
%
%
%
%
%
%
Integrating both sides of (\ref{current2}) over a closed loop inside a superconducting material (where $\vec J=0$),
we get:
\bea
0 = \oint \vec J \cdot d\vec l = \oint \le( \frac{2e^2}{mc}\vec A +  \frac{e\hbar}{m} \vec\nabla\phi\ri)\cdot d\vec l
\eea
or, from Stokes theorem:
\bea
\Phi &\equiv& \int \vec B \cdot d\vec S  = \oint \vec A\cdot d\vec l  =\frac{\hbar c}{2e}\oint \vec\nabla\phi \cdot d\vec l
\nn \\
&=&  \frac{\hbar c}{2e} \Delta\phi =\frac{\hbar c}{2e} 2\pi n  \equiv n \Phi_0~,~~\Phi_0 = \frac{ h c}{2e}~, n \in \mathbb{N}
\la{fq1}
\eea
which is the flux quantization in a superconductor.

Next, we would like to estimate GUP effects on the above flux quantum.
We see from (\ref{schr1}) that
because of the $\nabla^3$ operator, the leading Planck scale term of order $O(a)$ is non-local,  except in $1$-spatial
dimension. We do not know of a natural way of circumventing the problem for the non-relativistic case at hand, though such a linearization can modify the Dirac equation \cite{advplb1}  (which we shall use in Section \ref{muon}).
Thus we will work with the earlier version of the GUP and equivalently the $O(a^2)$ term in
Eq.(\ref{schr1}).
%
%
%
%
The new conserved current follows (see \cite{dvcjp}, with $\beta \rightarrow 5a^2/2$),
again for charge $-2e$ and mass $2m$
\bea
\vec J &=& \frac{\hbar}{2mi} \le[ \psi^\star \vec\nabla\psi - \psi\vec\nabla\psi^\star\ri]
\nn \\
&+& \frac{5a^2 \hbar^3 e }{2 mi}
\le[
\le(\psi^\star \vec\nabla \nabla^2 \psi - \psi \vec\nabla \nabla^2\psi^\star \ri) \ri.  \nn \\
&& \le. +\le( \nabla^2 \psi^\star \vec\nabla \psi - \nabla^2\psi \vec\nabla \psi^\star \ri)
\ri] \la{newcurrent1} \\
&\equiv& \vec{J}_0 + \vec{J}_1~, \la{newcurrent3} \\
\rho &=& |\psi|^2 \qquad
 \vec \nabla \cdot \vec J + \frac{\pa \rho}{\pa t} = 0~,
\eea
with $J_1$ being the GUP induced term. Once again, minimal coupling with the Cooper pairs give
\begin{widetext}
\bea
&&\vec J_1 =
-\frac{5a^2 e}{2m} \le\{
\psi^\star
\le[
\le( \frac{\hbar\vec\nabla}{i} + \frac{2e}{c}\vec A  \ri)
\le( \frac{\hbar\vec\nabla}{i} + \frac{2e}{c}\vec A  \ri) \cdot
\le( \frac{\hbar\vec\nabla}{i} + \frac{2e}{c}\vec A  \ri) \ri]
\psi
\ri.
  + \le[
\le( \frac{\hbar\vec\nabla}{i} + \frac{2e}{c}\vec A  \ri)
\le( \frac{\hbar\vec\nabla}{i} + \frac{2e}{c}\vec A  \ri) \cdot
\le( \frac{\hbar\vec\nabla}{i} + \frac{2e}{c}\vec A  \ri) \psi \ri]^\star
\psi \nn \\
&& +
\le[
\le( \frac{\hbar\vec\nabla}{i} + \frac{2e}{c}\vec A  \ri) \cdot
\le( \frac{\hbar\vec\nabla}{i} + \frac{2e}{c}\vec A  \ri)
 \psi \ri]^\star
\le[
\le( \frac{\hbar\vec\nabla}{i} + \frac{2e}{c}\vec A  \ri)
\psi \ri]
\le.
+
\le[
\le( \frac{\hbar\vec\nabla}{i} + \frac{2e}{c}\vec A  \ri) \cdot
\le( \frac{\hbar\vec\nabla}{i} + \frac{2e}{c}\vec A  \ri)
 \psi \ri]
 \le[
\le( \frac{\hbar\vec\nabla}{i} + \frac{2e}{c}\vec A  \ri)
\psi
\ri]^\star
\ri\}
\eea
\end{widetext}
Using $|\vec A| \approx |\vec B| L$, where $L$ is a typical linear dimension  of
the sample
and $\hbar \vec\nabla\psi/i = \hbar \psi \vec\nabla\phi \approx \hbar 2\pi \psi/L$,
an experiment can be arranged that
such that $|\hbar \vec\nabla\psi/i| \ll |2e\vec A\psi/c|$.
For example, for $|\vec B| \approx 0.1~T,~L \approx 0.1~m$,
$|\hbar \vec\nabla\psi/i|/ |2e\vec A\psi/c| \approx 10^{-3}$.
Hence
\be
\vec J_1 \approx -\frac{80a^2 e^4}{mc^3} \vec A |\vec A|^2 |\psi|^2~.
\ee
Thus using once again $0 = \oint \vec J\cdot d\vec l
=  \oint \vec J_0\cdot d\vec l + \oint \vec J_1 \cdot d\vec l  $, and treating
$|\vec A|^2 \approx  |\vec B|^2 L^2$ as effectively constant over the domain of
integration we now,  in lieu of Eq.(\ref{fq1}), get the flux
$(1 - \frac{40 a^2 e^2}{c^2} |\vec A|^2) \Phi = \frac{hc}{2e} n $ or
\bea
&& \Phi = \le(
1 + \frac{40 a^2 e^2}{c^2} |\vec A|^2
\ri) n \Phi_0 ~
\equiv n\le( \Phi_0 + a^2 \Phi_1 \ri)  \la{newphi} \\
&&
\Phi_1 = \frac{40~e^2 |\vec B|^2 L^2}{c^2}~\Phi_0
\la{phi1}~,~n\in \mathbb{N}~. \la{phi1}
\eea
to leading order in $a$.

Measurement of the fundamental flux quantum implies $a^2\Phi_1 < \d\Phi_0/\Phi_0$,  where $\d \Phi_0/\Phi_0$ is the experimental error.   Using Eq.(\ref{phi1}) above, we obtain
an upper bound on $a_0$,
\bea
a_0 &<& \frac{10^{-n/2}}{\sqrt{40}} \frac{M_{Pl}c^2}{eBL} <
 10^{19 -n/2}~, \la{a0bound}
\eea
assuming experimental precision of $1$ part in $10^n$,
where again we used $|\vec B| \approx 0.1~T,~L \approx 0.1~m$.
For example, for $n=4$, $a_0<10^{17}$. Conversely, if a significant improvement of precision
can be achieved, then small deviations from $\Phi_0$, as predicted above, may be observable!
 P induced correction to the flux quantum.

\section{Quantum Hall Effect}
\la{quantumhall}

The modification of the flux quantum has a direct effect on the observable
Hall resistance. As is well known, a current density $j_x$ along $x$ in a two dimensional sample in the $xy$ plane
subjected to a magnetic field $B$ along $z$ results in a potential difference and an effective electric field
${\cal E}_y$ along $y$.
This results in a cancelation of the electric and
Lorentz force on the charge carriers (having drift velocity $v$)
\bea
e{\cal E}_y = e vB ~,
\la{lorentz1}
\eea
and sets up a measurable potential difference in that direction.
Eq.(\ref{lorentz1}), and the relation $j_x=nev$, where $n$ is the electron density in the sample, imply
\bea
{\cal E}_y = \frac{j_x B}{ne}~.
\la{hall2}
\eea
The Hall resistivity $\rho_{xy}$ is defined by the relation
\bea
{\cal E}_y = \rho_{xy} j_x~,
\eea
which combined with Eq.(\ref{hall2}) yields
\bea
\rho_{xy} = \frac{B}{ne}~.
\la{hallresist}
\eea

We also know that quantum mechanically,
the electrons in the sample subjected to the perpendicular magnetic field give rise to
Landau levels, with the energy at level $n$ given by
\bea
E_n = \hbar \omega_c \le( n+ \frac{1}{2} \ri)~,
\eea
where $\omega_c=eB/mc$ is the cyclotron frequency. Now, from
flux quantization it follows that the density of quanta of magnetic flux is given by
%
\bea
n_H = \frac{B}{\Phi_0 + a^2 \Phi_1}~
\eea
in terms of the single unit of flux quanta given by (\ref{newphi}).
This is also the density of states for each Landau level,
where we have made the replacement $2e \rightarrow e$ in Eq.(\ref{fq1}),
since now the carriers are electrons as opposed to Cooper pairs.

Now if the Fermi energy $E_F$ lies between the energy levels $E_k$ and $E_{k+1}$,
all states $E_i~,i\leq k$ are occupied, resulting in the carrier density
\bea
n = k n_H,~ 
k\in \mathbb{N} .
\eea
Thus from Eqs.(\ref{newphi}), (\ref{phi1}) and (\ref{hallresist}), the Hall resistance turns out to be
\cite{hallweb}
\bea
\rho_{xy} = \frac{hc}{ke^2}
\le[ 1 + \frac{10 a^2 e^2 |\vec B|^2 L^2 }{c^2} \ri]~, 
k\in \mathbb{N}~.
\eea
Although the Hall resistance is still quantized, its magnitude has shifted by a small amount.
A bound similar to Eq.(\ref{a0bound}), as well as possibilities of
measurement of corrections of the above type can be argued for this case as well.

%

\section{Anomalous magnetic moment of the muon}
\la{muon}
In this case, we show that the non-relativistic limit of the Dirac equation can be
used to extract GUP corrections.
First, as in \cite{advplb1}
we linearize
$p_0 = \sqrt{p_{0x}^2+p_{0y}^2+p_{0z}^2}$ by replacing
$p_0 \rightarrow \vec\alpha \cdot \vec p$,
where $\alpha_i~(i=1,2,3)$ and $\b$ are the Dirac matrices, for which we
use the following representation
\bea
\a_i =
\left( \begin{array}{cc}
0 & \sigma_i \\
\sigma_i & 0 \end{array} \right)~,~
\b =
\left( \begin{array}{cc}
I & 0 \\
0 & -I \end{array} \right)~.
\eea
The GUP-corrected Dirac equation can thus be written to ${\cal O}(a)$ as
\bea
H \psi &=& \le (c\, \vec \a \cdot \vec p + \b mc^2 \ri) \psi (\vec r,t)  \nn \\
&=&  \le(c\, \vec \a \cdot \vec p_0 -
c\, a (\vec\a \cdot \vec p_0)(\vec\a \cdot \vec p_0) + \b mc^2 \ri) \psi (\vec r,t)\nn \\
&=& i\hbar \frac{\pa \psi (\vec r,t)}{\pa t}~.
\la{ham1}
\eea
To study the non-relativistic limit, we write the spinor $\psi$ as
\cite{mannbook}
\bea
\psi &=& e^{-imt}\left( \begin{array}{c}
\chi_1 (\vec r,t) \\
\chi_2 (\vec r,t) \end{array} \right) ~,
\eea
and we
include the electromagnetic potential $A^\mu=(\phi,\vec A)$ in Eq.(\ref{ham1})
by the usual minimal coupling prescription  \cite{dvcjp}
$ i\hbar \frac{\pa}{\pa t} \rightarrow i\hbar \frac{\pa}{\pa t} - e\phi,~
\vec p_0 \rightarrow \vec\Pi_0 \equiv \vec p_0 - e\vec A/c$,
obtaining the two component equations
\bea
i\hbar \frac{\pa \chi_1}{\pa t} &=& e \phi \chi_1 + c\le( \vec\sigma\cdot\vec\Pi \ri) \chi_2
-ca \le( \vec\sigma\cdot\vec \Pi\ri)^2 \chi_1
\la{chi1} \\
i\hbar \frac{\pa \chi_2}{\pa t} &=& \le( e \phi -2mc^2 \ri) \chi_2  + c \le( \vec\sigma\cdot\vec\Pi \ri) \chi_1
-ca \le( \vec\sigma\cdot\vec \Pi\ri)^2 \chi_2  \nn
\eea
In the non-relativistic limit $mc^2 >> e\phi, |\pa \chi_2/\pa t|$, the second of Eqs.(\ref{chi1}) becomes to ${\cal O}(a)$
\bea
\chi_2
=
\frac{1}{2mc}
\le[1  - \frac{a}{2mc} \le( \vec\sigma\cdot\vec\Pi \ri)^2  \ri]
\le( \vec\sigma\cdot\vec\Pi \ri) \chi_1
~,
\eea
which, when substituted into the first of Eqs.(\ref{chi1}) yields
\bea
&& i\hbar \frac{\pa \chi_1}{\pa t} =
e \phi \chi_1 + \frac{1}{2m}\le( \vec\sigma\cdot\vec\Pi \ri)^2 \chi_1 \nn \\
&-&  \frac{a}{(2m)^2c} \le( \vec\sigma\cdot\vec\Pi \ri)^4
\chi_1 -  ca \le( \vec\sigma\cdot\vec\Pi \ri)^2 \chi_1~.
\la{chi2b}
\eea
Using the identities $\sigma_a \sigma_b = \delta_{ab} + i\epsilon_{abc}\sigma_c$
and  $\le( \vec\sigma\cdot\vec\Pi \ri)^2=|\vec \Pi|^2 - e \hbar \vec \sigma \cdot \vec B/c$
and the identification of the spin operator $\vec S = \vec \sigma/2$,
Eq.(\ref{chi2b}) becomes
\bea
&&  i\hbar \frac{\pa \chi_1}{\pa t} =
\le[
\le( \frac{1}{2m} -ca \ri) |\vec\Pi|^2
-\frac{a}{(2m)^2c} \Pi^4
+ e \phi \chi_1
\ri.
\nn \\
&&
- 2\frac{e\hbar}{2mc} \le( 1 - 2acm - \frac{a}{mc} \Pi^2  \ri) \vec S \cdot \vec B
- \frac{ae^2\hbar^2}{(2m)^2c^3} |\vec B|^2
 \\
&& \le.
- \frac{ie\hbar a}{2(mc)^2}
\le( \vec\nabla (\vec\sigma\cdot\vec B) \cdot \vec\Pi
- \frac{ e}{c} \vec A \cdot \vec\nabla (\vec\sigma\cdot\vec B)
+i\hbar \nabla^2(\vec\sigma\cdot\vec B)
\ri)
\ri]
 \chi_1 \nn
\eea
where the terms in the first line correspond to the GUP corrected kinetic terms
(including the $\Pi^4$ term) and the potential energy term, while
those on the second and third lines pertain to the interaction with the electron with an
external magnetic field. The ones in the third line are also new terms
which depend on derivatives of the magnetic field\footnote{We used the following identities in their derivation:
$[\vec\Pi,\vec\sigma\cdot\vec B]=\vec p_0 (\vec \sigma \cdot\vec B)$,
$[\Pi^2,\vec\sigma\cdot\vec B]=\vec\Pi\cdot\vec p_0(\vec\sigma\cdot\vec B) +
\{ \vec p_0 (\vec\sigma\cdot\vec B)\}\cdot\vec \Pi$ and
$(\vec \sigma \cdot \vec B) \Pi^2 = \Pi^2 (\vec \sigma \cdot \vec B) - [\Pi^2,\vec\sigma\cdot\vec B] $
}
.
Since $e/2mc$ is the Bohr magneton, one gets
$
g = 2 (1 - 2acm - \frac{a}{mc} \Pi^2)~,
$
or
\bea
\le( \frac{g-2}{2}\ri)_{GUP} = - \le[2acm + \frac{a\Pi^2}{mc}  \ri] ~.
\eea
%
%
%
Note that GUP predicts a slight decrease in the value of $g$, and for a measurement accuracy of
$1$ part in $10^n$, one obtains the bound
\bea
a_0 &<& 10^{-n} \frac{m_{Pl}}{m_\mu} \nn \\
&<& 10^{20-n}~,
\eea
where we have used $m_{Pl}=1.2 \times 10^{19}~GeV/c^2$ and $m_\mu=105.7~MeV/c^2$.
Thus for the present-day precision level of $n=12$, $a_0 < 10^8$, which gives a much tighter bound. Conversely,
further increase of accuracies may enable one to observe the above deviation.

\section{Conclusions}

In this paper we have explored Planck scale effects on some low energy systems via the Generalized
Uncertainty principle, which appears to be a robust prediction of most theories of Quantum
Gravity.
We found that small but non-zero effect are present for the fundamental flux quantum of
superconductivity, for the integer quantum Hall effect, and for the anomalous magnetic moment of the muon.
Since these effects have not been observed so far, one can obtain important upper bounds on the
GUP parameter, which turns out to be $a_0<10^{17}$ and $a_0<10^8$ from current experiments in
Superconductivity and muon experiments respectively. These can be compared with the bounds obtained
in \cite{dvprl}, between $10^{18}$ and $10^{25}$, and in \cite{advprd}, between
$10^{10}$ and $10^{23}$.
Although the above bounds appear to be rather weak,
future experiments of greater precision  will either provide better bounds on the GUP parameter or
in an optimistic scenario, may be able to detect some of these effects.
In a broader context this approach may open up a low energy window to quantum
gravity phenomenology. This strongly suggests that more work needs to be done in this direction.


\begin{center}
{\bf Acknowledgment}
\end{center}

We thank the anonymous referee for useful comments which helped improve the paper.
This work was supported in part by the Natural
Sciences and Engineering Research Council of Canada and by the
Perimeter Institute for Theoretical Physics, Waterloo, Canada,
where part of the work was done.

\vs{.3cm}
\no
{\bf Note added} \\
After completion of this work, we became aware of paper \cite{hossenfeldergminus2}
(we thank the referee for pointing this out to us), in which the authors
obtain a lower bound on the fundamental (higher dimensional) Planck mass in
theories with extra dimensions, from muon $(g-2)$ measurements,
by directly using the modified Dirac equation.
In this paper on the other hand, we use the non-relativistic
limit of the GUP modified Dirac equation to obtain upper bounds on $a_0$.




\begin{thebibliography}{99}

\bibitem{guppapers} D. Amati, M. Ciafaloni, G. Veneziano,
Phys. Lett. B {\bf 216} (1989) 41;
M.~Maggiore,
 Phys.\ Lett.\  B {\bf 304} (1993) 65  [arXiv:hep-th/9301067];
M.~Maggiore,
  Phys.\ Rev.\  D {\bf 49} (1994) 5182 [arXiv:hep-th/9305163];
M.~Maggiore,
  Phys.\ Lett.\  B {\bf 319} (1993) 83
  [arXiv:hep-th/9309034];
L.~J.~Garay, Int.\ J.\ Mod.\ Phys.\  A {\bf 10} (1995) 145 [arXiv:gr-qc/9403008];
%
F.~Scardigli,
  Phys.\ Lett.\  B {\bf 452} (1999) 39
  [arXiv:hep-th/9904025];
  %
  %
S.~Hossenfelder, M.~Bleicher, S.~Hofmann, J.~Ruppert, S.~Scherer and H.~Stoecker,
  Phys.\ Lett.\  B {\bf 575} (2003) 85
  [arXiv:hep-th/0305262];
  %
  %
C.~Bambi and F.~R.~Urban,
  Class.\ Quant.\ Grav.\  {\bf 25} (2008) 095006
  [arXiv:0709.1965 [gr-qc]].

\bibitem{kmm} A. Kempf, G. Mangano, R. B. Mann, Phys. Rev. D {\bf
 52} (1995) 1108 [arXiv:hep-th/9412167].

\bibitem{kempf} A. Kempf, J.Phys. A  {\bf 30} (1997) 2093 [arXiv:hep-th/9604045].

\bibitem{brau} F. Brau, J. Phys. A {\bf  32} (1999) 7691 [arXiv:quant-ph/9905033].

\bibitem{sm} J.~Magueijo and L.~Smolin,
  Phys.\ Rev.\ Lett.\  {\bf 88} (2002) 190403
  [arXiv:hep-th/0112090];
J.~Magueijo and L.~Smolin,
  Phys.\ Rev.\  D {\bf 71} (2005) 026010
  [arXiv:hep-th/0401087].

\bibitem{cg}
J. L. Cortes, J. Gamboa, Phys. Rev. D {\bf 71} (2005) 065015 [arXiv:hep-th/0405285];

\bibitem{viqar}
G. M. Hossain, V. Husain, S. S. Seahra,
arXiv: 1003.2207 [gr-c].

\bibitem{dvprl} S. Das, E. C. Vagenas, Phys. Rev. Lett. {\bf 101} (2008) 221301 [arXiv:0810.5333 [hep-th]].

\bibitem{advplb1} A. Ali, S. Das, E. C. Vagenas,
Phys. Lett. {\bf B678} (2009) 497-499 [arXiv:0906.5396].

\bibitem{advprd}
A. Ali, S. Das, E. C. Vagenas,
Phys. Rev. {\bf D84} (2011) 044013 [arXiv:1107.3164 [hep-th]].
See Appendix A for proof of Eq.(\ref{comm01}).

\bibitem{dvcjp} S. Das, E. C. Vagenas,
Can. J. Phys. {\bf 87} (2009) 233 [arXiv:0901.1768 [hep-th]].


\bibitem{advplb2} S. Das, E. C. Vagenas, A. Ali,
Phys. Lett. {\bf B690} (2010) 407-412 [arXiv:1005.3368].

\bibitem{am}
N. W. Ashcroft, N. D. Mermin,
{\it Solid State Physics},
Brooks/Cole (1976).


\bibitem{hallweb}
http:\/\/www.nonequilibrium.net\/191-integer-quantum-hall-effect-theory\/

\bibitem{mannbook}
R. B. Mann, {\it An Introduction to Particle Physics and the Standard Model},
CRC Press (2009), Chapter 14.

\bibitem{hossenfeldergminus2}
U. Harbach, S. Hossenfelder, M. Bleicher, H. St\"ocker,
Phys. Lett. {\bf B584} (2004) 109 [arXiv:hep-ph/0308138]



\end{thebibliography}
\end{document}